\documentclass[journal]{IEEEtran}
\ifCLASSINFOpdf
   \usepackage[pdftex]{graphicx}
   \DeclareGraphicsExtensions{.pdf,.jpeg,.png,.eps}
\else

   \usepackage[dvips]{graphicx}
\fi
\usepackage[cmex10]{amsmath}

\usepackage{array}
\usepackage{mdwmath}
\usepackage{mdwtab}
\usepackage{eqparbox}
\usepackage{adjustbox}
\usepackage{booktabs}
\usepackage{color}

\usepackage[tight,footnotesize]{subfigure}
\usepackage{multirow}
\usepackage{cite}
\usepackage{stfloats}

\begin{document}
\title{Analytical Modeling and Design of Gallium Oxide Schottky Barrier Diodes Beyond Unipolar Figure of Merit Using High-k Dielectric Superjunction Structures}
 
\author{Saurav~Roy,
        ~Arkka Bhattacharyya,
        ~and Sriram Krishnamoorthy      

 \vspace{-0.42cm}
\thanks{Saurav Roy, Arkka Bhattacharyya, and Sriram Krishnamoorthy are with the Department of Electrical and Computer Engineering, The University of Utah, Salt Lake City, UT, 84112, United States of America (e-mail: saurav.roy@utah.edu; a.bhattacharyya@utah.edu; sriram.krishnamoorthy@utah.edu).}}


\maketitle
\begin{abstract}
This work presents the design of $\beta$-Ga$_2$O$_3$ schottky barrier diode using high-k dielectric superjunction to significantly enhance the breakdown voltage vs on-resistance trade-off beyond its already high unipolar figure of merit. The device parameters are optimized using both TCAD simulations and analytical modeling using conformal mapping technique. The dielectric superjunction structure is found to be highly sensitive to the device dimensions and the dielectric constant of the insulator. The aspect ratio, which is the ratio of the length to the width of the drift region, is found to be the most important parameter in designing the structure and the proposed approach only works for aspect ratio much greater than one. The width of the dielectric layer and the dielectric constant also plays a crucial role in improving the device properties and are optimized to achieve maximum figure of merit. Using the optimized structure with an aspect ratio of 10 and a dielectric constant of 300, the structure is predicted to surpass the $\beta$-Ga$_2$O$_3$ unipolar figure of merit by four times indicating the promise of such structures for exceptional FOM vertical power electronics.
\end{abstract}

\maketitle

%
\vspace{-0.1cm}
$\beta$-Ga$_2$O$_3$ has garnered a lot of attention in recent years for power device applications due to its high breakdown
field. $\beta$-Ga$_2$O$_3$ has a band gap (4.6 eV) larger than GaN and SiC, with an estimated critical breakdown field as high as 8 MV/cm. Due to the large critical electric field, the Baliga Figure of Merit (BFOM) relevant to power switching could be 2000–3400 times that of Si, which is several times larger than that of SiC or GaN. Low doped drift layers in conjunction with large band gap materials can enable very high breakdown voltage. Various power devices using $\beta$-Ga$_2$O$_3$ have been demonstrated recently with high breakdown voltage in the vertical and lateral geometry \cite{lin2019,zhou2019,gao2019high,konishi20171,yang2017high,yang20182300v,li20182,li20181230,uttam,yang2017highb,xia2019,chandan, yuhao,kelson}. 

Superjunction techniques using complementary doped columns of silicon have been employed in silicon power devices to improve trade-off between breakdown voltage and on resistance which enabled the use of silicon based devices at larger voltage rating power applications despite of its low FOM \cite{onishi, deboy, kurosaki, yamauchi}. In conventional superjunction structures multiple p and n pillars are used and the doping of those layers controls the breakdown voltage enhancement in those devices. In these structures, due to the charge sharing between the p and n layers, a lateral electric field component is induced. Now, because there are two components of electric field, 1) vertical component due to applied bias (anode to cathode) and 2) lateral field due to the space charge coupling between p and n layers, the overall field profile changes from a triangular shape to a rectangular one, which is the key to attaining higher breakdown down voltages. However a very narrow range of doping is allowed in superjunction structures to achieve maximum performance which poses fabrication difficulties in making these devices \cite{udrea, shenoy}. In $\beta$-Ga$_2$O$_3$ based devices, the lack of p-type doping makes the realization of conventional superjunction structures extremely challenging. Dielectric resurf technique \cite{sonsky} can be used in this case to circumvent the issue of lack of shallow acceptors in gallium oxide. High-k dielectrics have been recently explored for electric field management in wide band gap semiconductors-based lateral and vertical device structures \cite{nidhin,cheng,razzak2020batio3,xia2019}. If high-k dielectric is used instead of a p-type layer, the difference in the dielectric constants between the drift layer and the insulator will create a lateral electric field, which can flatten the electric field as in the case of conventional superjunction structures. Furthermore due the wider band-gap of $\beta$-Ga$_2$O$_3$, creating dielectric superjunction structure with reasonable aspect ratio for enhanced breakdown is more feasible than its smaller bandgap counterparts \cite{sabui}. However very careful designing is necessary to achieve a performance boost as it is very sensitive to the structure dimensions.

Vertical dielectric resurf structures have been modeled previously by several groups \cite{zhou1,sabui} considering multiple fingers. In these models Poisson's equation was solved considering symmetry lines at the dielectric boundary. However, the structure is found to be highly dependent on the aspect ratio (A.R), which is the ratio of the length of the drift layer (Anode to Cathode) to the width of the same and the performance only improves if the aspect ratio is greater than one. This high aspect ratio requirement makes the fabrication of vertical superjunction structures very challenging. So a lateral structure as shown in Fig. 1(b) would be more feasible to fabricate. But in a lateral structure the absence of symmetry will cause the model presented by zhou et al. \cite{zhou1} to fail. In such asymmetric structures the fringing electric field through the dielectric should be modeled appropriately to get accurate results.

In this work an analytical model for the structure has been developed by solving the Poisson's equation in both the drift layer and in the dielectric. The fringing electric field through the dielectric layer has been modeled using conformal mapping which makes the model more accurate in describing both symmetric and non-symmetric structures. The optimum dimensions were determined from the model and also been verified with numerical TCAD simulations using synopsis Sentaurus \cite{sentaurus}. A very good agreement between the model and the TCAD results demonstrate the accuracy of the model. 

\begin{figure*}[t]
\centering
\subfigure{
\includegraphics[width=1.5in, height=1.4in, keepaspectratio]{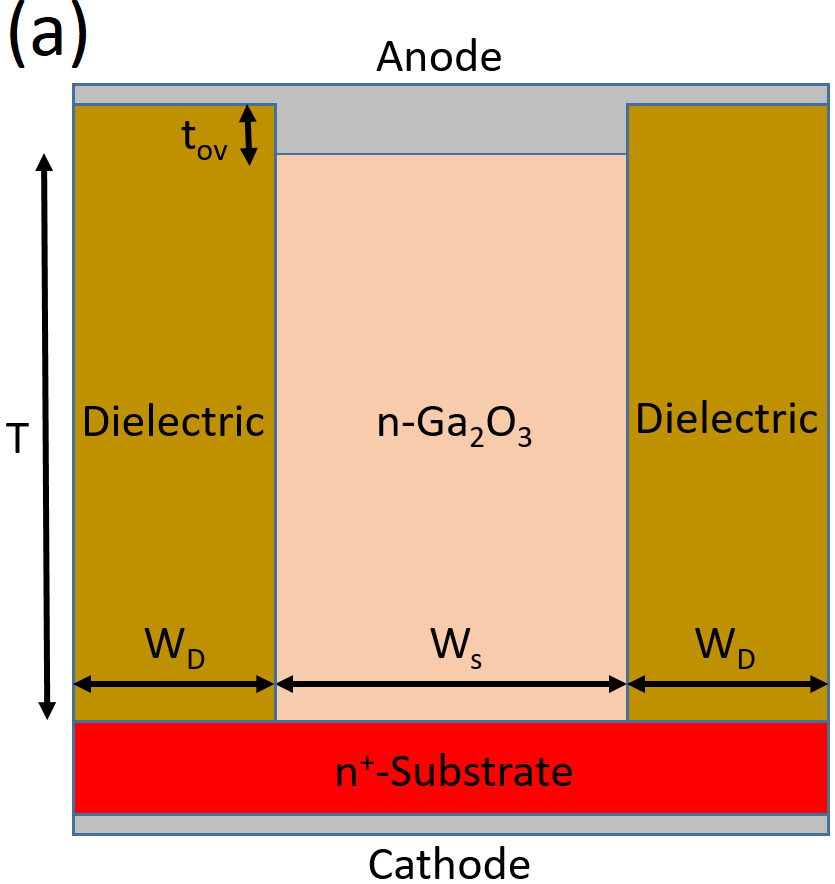}}\hspace{1cm}
\subfigure{
\includegraphics[width=1.5in, height=1.4in, keepaspectratio]{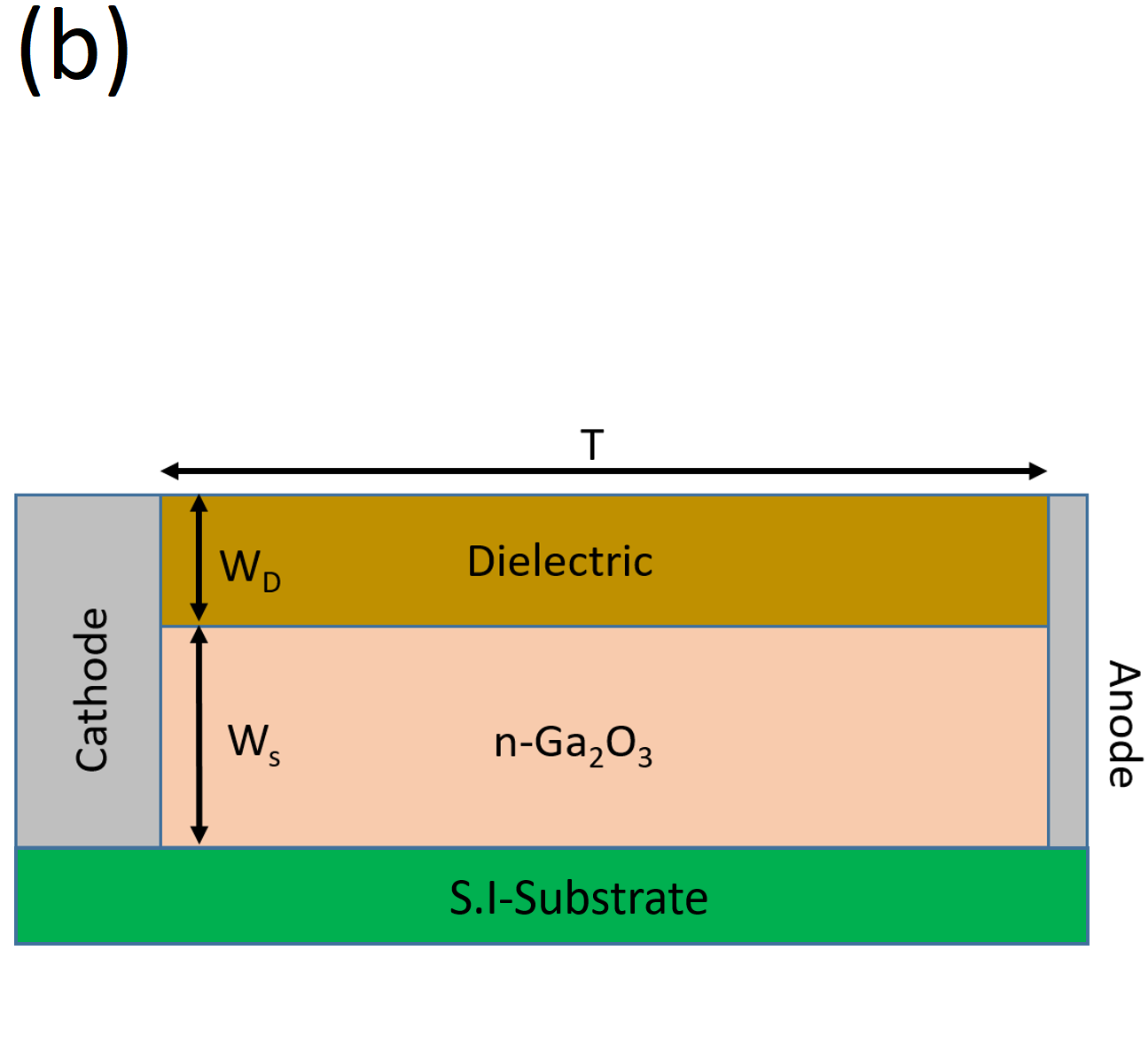}}\hspace{-0.2cm}
\subfigure{
\includegraphics[width=1.5in, height=1.4in, keepaspectratio]{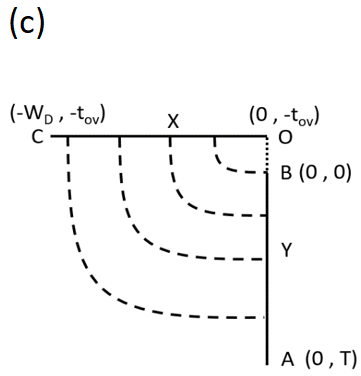}}\hspace{-0.2cm}
\subfigure{
\includegraphics[width=2.1in, height=2.5in,keepaspectratio]{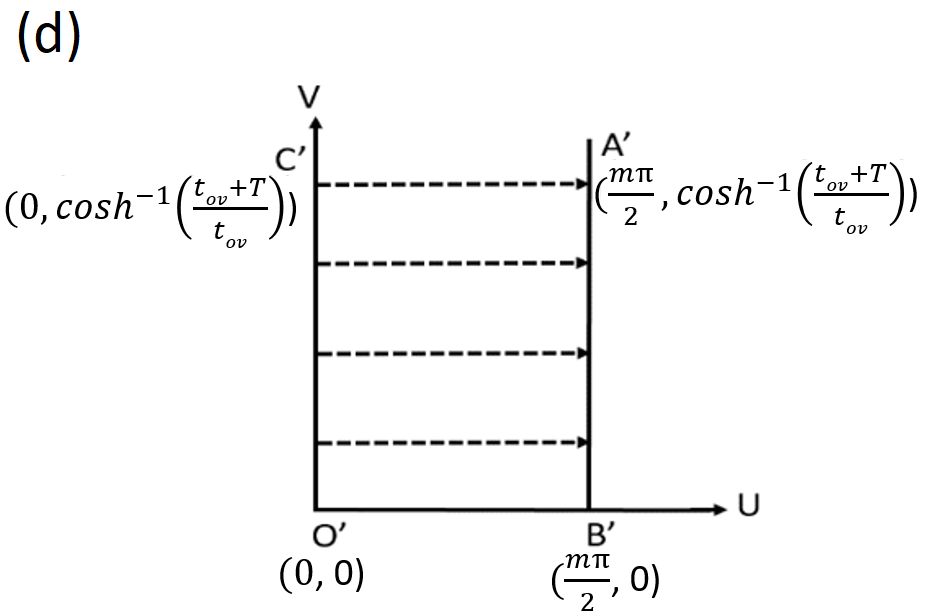}}\hspace{-0.2cm}
\caption{\footnotesize Schematic diagram of (a) vertical (b) lateral dielectric superjunction SBD. (c) Conformal mapping transformation of left side of the device. AB is the semiconductor oxide interface, OC is the top edge of the oxide. (b) XY coordinate is transformed to UV coordinate. }
\label{fig2}
\vspace{-0.2cm}
\end{figure*}

The 2-D Poisson equation of a dielectric superjunction structure with doped gallium oxide drift region as shown in Fig. 1(a) can be written as,

\begin{equation}
\label{eq1}
\frac{\partial^2\psi (x,y)}{\partial x^2} + \frac{\partial^2\psi (x,y)}{\partial y^2} = - \frac{\rho}{\epsilon}
\end{equation}

where,
 \[
   \rho= 
\begin{cases}
    qN_D,           &  x> 1\\
    0,              & x<1
\end{cases}
\]

where,  $\psi (x,y)$ is the electrostatic potential at the channel region, q represents the electron charge, N$_D$ is the doping concentration inside the drift region.

The boundary conditions at top and bottom contacts are considered as,
\begin{equation}
   \psi(x,0) = -V_R 
\end{equation}

\vspace{-0.7cm}

\begin{equation}
    \psi(x,T) = 0
\end{equation}

where V$_R$ is the applied reverse bias voltage. Now, if the structure is repetitive with multiple pillars of dielectric and semiconductor region then using symmetry of the structure, the electric field at the dielectric boundary can be considered as zero. However, if an asymmetric structure is present with only one semiconductor and dielectric pillar then the fringing electric field through the dielectric has to be taken into account. 

Neumann boundary condition of continuity of electric displacement can be applied at the dielectric semiconductor interface considering elliptical field lines emanating from the anode as shown in Fig. 1(c). These fringing fields can be modeled using conformal mapping technique \cite{plonsey1961electromagnetic,bansal} which maps the curved geometry to a straight geometry as shown in Fig. 1(d). An appropriate transfer function to model this case can be written as;
\vspace{-0.1cm}
\begin{equation}
\label{eq9}
(y+t_{ov}) + j\gamma x= t_{ov}sin(u+jv)
\end{equation}
where
$$
\gamma = \frac{t_{ov}}{W_D}sinh\left[cosh^{-1}\left(\frac{t_{ov} + T}{t_{ov}}\right)\right]
$$

where t$_{ov}$, W$_D$, and T are the dielectric overlap length and dielectric width and length of the drift region respectively as shown in Fig. 1(a).

By using the above transformation $ABOC$ of Fig. 1(c) in X-Y coordinate is transformed into $A'B'O'C'$ of Fig. 1(d) in U-V coordinate. By equating real and imaginary part of (\ref{eq9}) we can find the values of $x$ and $y$ in terms of $u$ and $v$ as shown below.
\vspace{-0.1cm}
\begin{align}
\label{eq10.1}
x=&\left(\frac{t_{ov}}{\gamma}\right)cos(u)sinh(v) \\ y=&-t_{ov}+t_{ov}sin(u)cosh(v)
\end{align}

We can map the coordinates of $O'$, $B'$, $C'$, and $A'$  from the above values of $x$ and $y$ as shown in Fig. 1(c) and 1(d).

From this transformation, the fringing capacitance is calculated to be,
\vspace{-0.1cm}
\begin{equation}
\label{eq10}
C_{fr,ox} = \frac{2\epsilon_{ox}}{m\pi}cosh^{-1}\left(\frac{t_{ov}+T}{t_{ov}}\right)
\end{equation}

Similarly the fringing capacitance in the semiconductor side can be written as,

\vspace{-0.1cm}
\begin{equation}
\label{eq10}
C_{fr,S} = \frac{2\epsilon_{S}}{m\pi}cosh^{-1}\left(\frac{t_{ov}+T}{t_{ov}}\right)
\end{equation}

The value of '$m$' is taken to be 5 by fitting the model with numerical simulation. Using all the boundary conditions mentioned above the expression for $\psi (x,y)$ can be written as, 
\vspace{-0.1cm}
\begin{multline}
\label{eq23}
\psi(x,y) = -\frac{qN_Dy^2}{2\epsilon_S}+ \left(\frac{qN_DT}{2\epsilon_S} -\frac{V_R}{T}\right)y +V_R + \\ \sum_{n=1}^{\infty} R_nsin(k_ny)\left[F_ne^{k_nx} +e^{-k_nx}\right]
\end{multline}

where,
\begin{equation}
\label{eq28}
k_n = \frac{n\pi}{W}
\end{equation}
\vspace{-0.3cm}
\begin{equation}
\label{eq29}
R_n = \lambda_n\frac{\epsilon_{ox}(F_n' - 1)}{(F_nF_n'-1)(\epsilon_{ox} -\epsilon_S)+(F_n'-F_n)(\epsilon_{ox} + \epsilon_S)}
\end{equation}
\vspace{-0.3cm}
\begin{equation}
   F_n = \frac{k_n\epsilon_SW_D + C_{fr,ox}}{k_n\epsilon_SW_D - C_{fr,ox}}
\end{equation}
\vspace{-0.3cm}
\begin{equation}
  F_n' = \frac{k_n\epsilon_{ox}W_S + C_{fr,S}}{k_n\epsilon_{ox}W_S - C_{fr,S}}
\end{equation}
\vspace{-0.1cm}
and 
\begin{equation}
\lambda_n = \left(\frac{qN_DT^2}{2\epsilon_S}\right)\times \frac{4}{(n\pi)^3}\left[(-1)^n - 1\right]
\end{equation}

The electric field has been calculated through the gradient of $\psi(x,y)$. The maximum electric field is located at the point ($\frac{W_S}{2}$,0) as shown by circle in the contour plots of Fig. 2 (a) and (b) and can be expressed as,
\begin{multline}
\label{eq30}
E_{max} = E(W_S/2,0) \\ = \sum_{n=1}^{\infty} R_nk_n\left[F_ne^{k_nW_S/2} +e^{-k_nW_S/2}\right] \\ + \left(\frac{qN_DT}{2\epsilon_S} -\frac{V_R}{T}\right)
\end{multline}

\begin{figure}[t]
\centering
\subfigure{
\includegraphics[width=1.71in, height=1.35in,keepaspectratio]{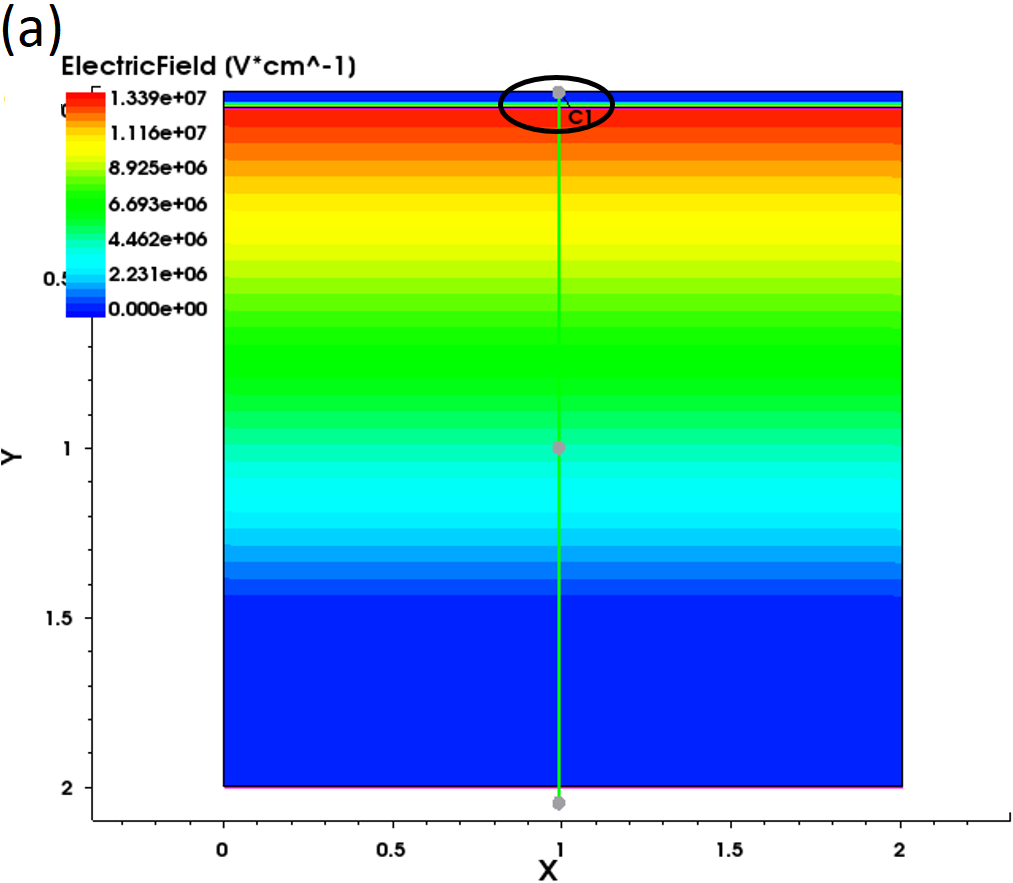}}\hspace{-0.2cm}
\subfigure{
\includegraphics[width=1.71in, height=1.35in,keepaspectratio]{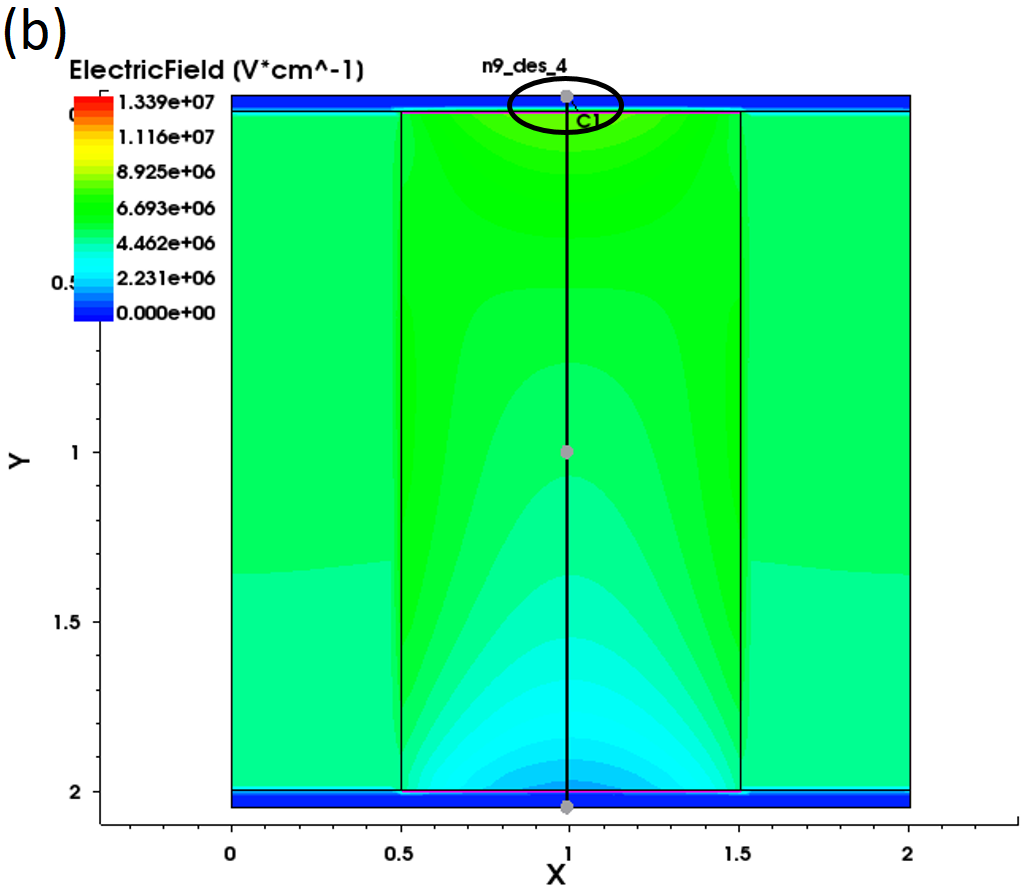}}
\\
\subfigure{
\includegraphics[width=2in, height=2in,keepaspectratio]{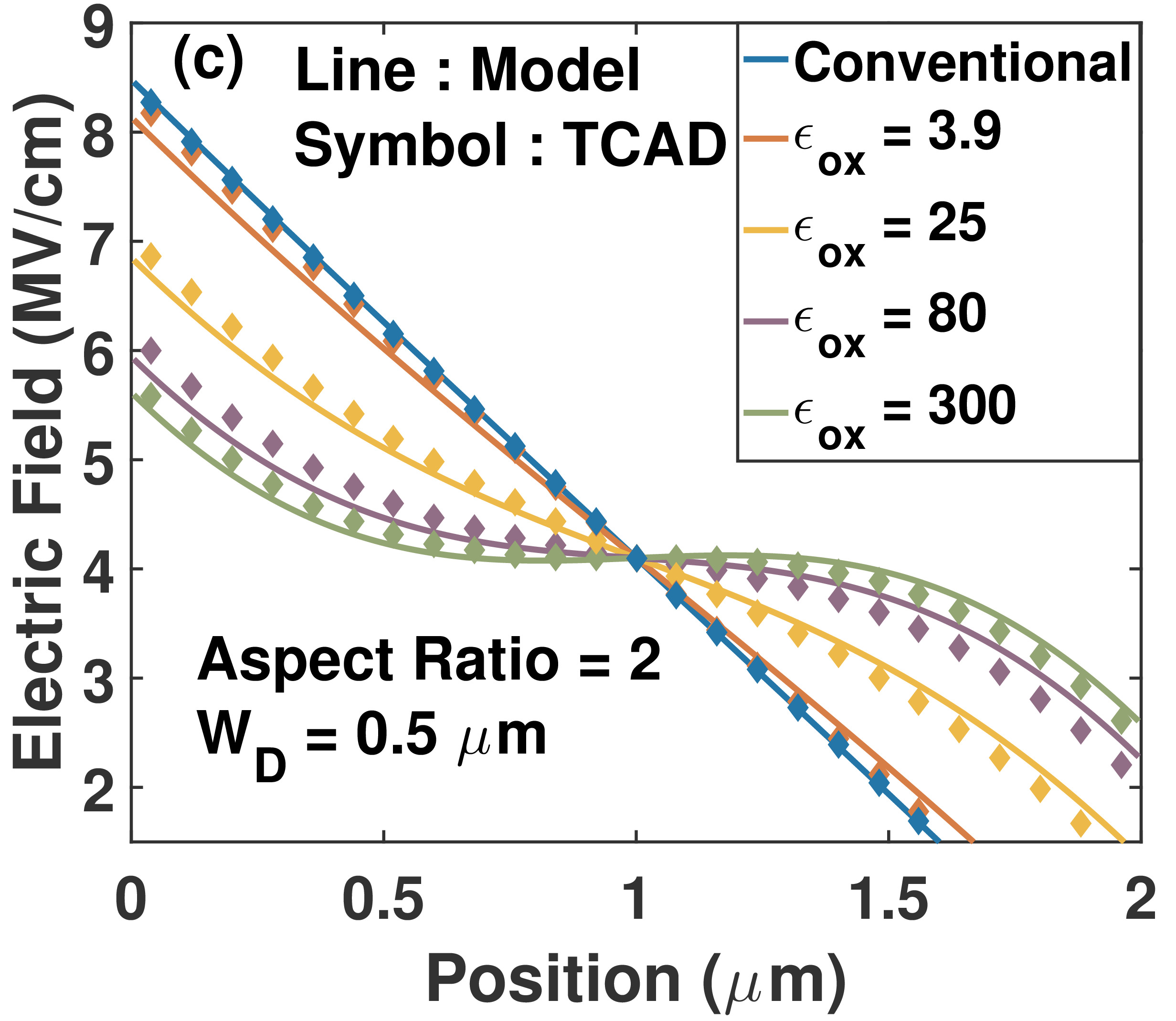}}
\caption{\footnotesize Electric field contour for (a) conventional SBD and (b) DS SBD structure with a dielectric material of dielectric constant 300 (c) Electric field distribution inside the DS SBD along the cutlines shown for aspect ratio greater than 1}
\label{fig7}
\vspace{-0.2cm}
\end{figure}

Fig. 2 shows the distribution of the magnitude of the electric field from the analytical model and compares it with the 2D numerical simulation of a dielectric modulated schottky diode using Sentaurus TCAD \cite{sentaurus}. From this figure we can see that as the dielectric constant of the dielectric material increases, the electric field profile changes from a triangular shape to a rectangular profile. For this work we have considered hypothetical dielectric materials. Nevertheless, there are several prospective high-k material choices, especially perovskites which offer such high dielectric constants \cite{xia}.

Breakdown voltage has been estimated as the reverse voltage at which the the maximum electric field reaches the theoretical critical electric field of $\beta$-Ga$_2$O$_3$ (\textasciitilde 8MV/cm). Thus equating E$_{max}$ in (\ref{eq30}) to E$_C$, we can get the breakdown voltage V$_{BR}$ as,

\vspace{-0.5cm}

\begin{multline}
\label{eq31}
V_{BR} = \left.V_R\right|_{E_{max} = E_C} \\ = \sum_{n=1}^{\infty} TR_nk_n\left[F_ne^{k_nW_S/2} +e^{-k_nW_S/2}\right] \\ + \left(\frac{qN_DT^2}{2\epsilon_S} -E_CT\right)
\end{multline}

\vspace{-0.0cm}
Now specific on resistance for this structure can be expressed as,

\vspace{-0.1cm}

\begin{equation}
\label{eq32}
R_{on} = \frac{T}{q\mu_nN_D}\left(1+\frac{W_D}{W_S}\right)
\end{equation}

where $\mu_n$ is the electron mobility. We can combine (\ref{eq31}) and (\ref{eq32}) to get the breakdown voltage in terms of R$_{on}$ as,

\begin{multline}
\label{eq33}
V_{BR} = \sum_{n=1}^{\infty} \frac{q\mu_nN_DR_{on}}{1+\frac{W_D}{W_S}}R_nk_n\left[F_ne^{k_nW_S/2} +e^{-k_nW_S/2}\right] \\ + \left(\frac{q^3N_D^3\mu_n^2R_{on}^2}{2\epsilon_S\left(1+\frac{W_D}{W_S}\right)^2} -E_C\frac{q\mu_nN_DR_{on}}{1+\frac{W_D}{W_S}}\right)
\end{multline}

The breakdown voltage is plotted against device parameters and to confirm the accuracy, it has been compared with the results acquired from the Sentaurus device simulator. The material parameters and physics used in both the simulation and model are provided in the supplementary material. The device breakdown voltage is extracted from E-field simulations when the peak E-field reaches the $\beta$-Ga$_2$O$_3$ (8 MV/cm) critical E-field. The ionization integrals for avalanche breakdown were not evaluated in order to avoid excessive computation time. Furthermore, accurate experimentally verified ionization rate parameters are currently unknown for $\beta$-Ga$_2$O$_3$.

                                                             
                                                             


\begin{figure}[t]
\centering
\subfigure{
\includegraphics[width=1.7in, height=1.4in,keepaspectratio]{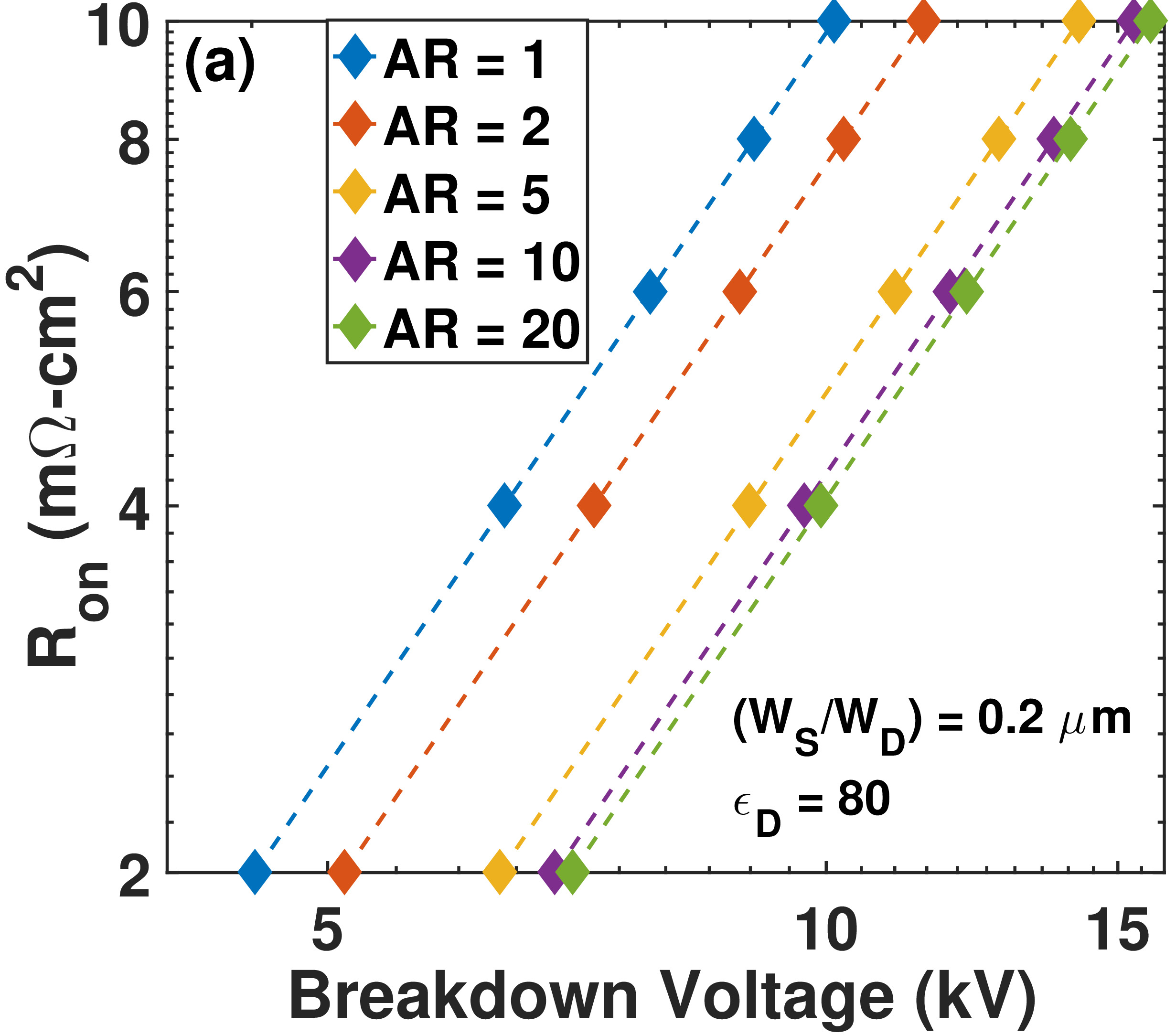}}\hspace{0cm}
\subfigure{
\includegraphics[width=1.7in, height=1.4in,keepaspectratio]{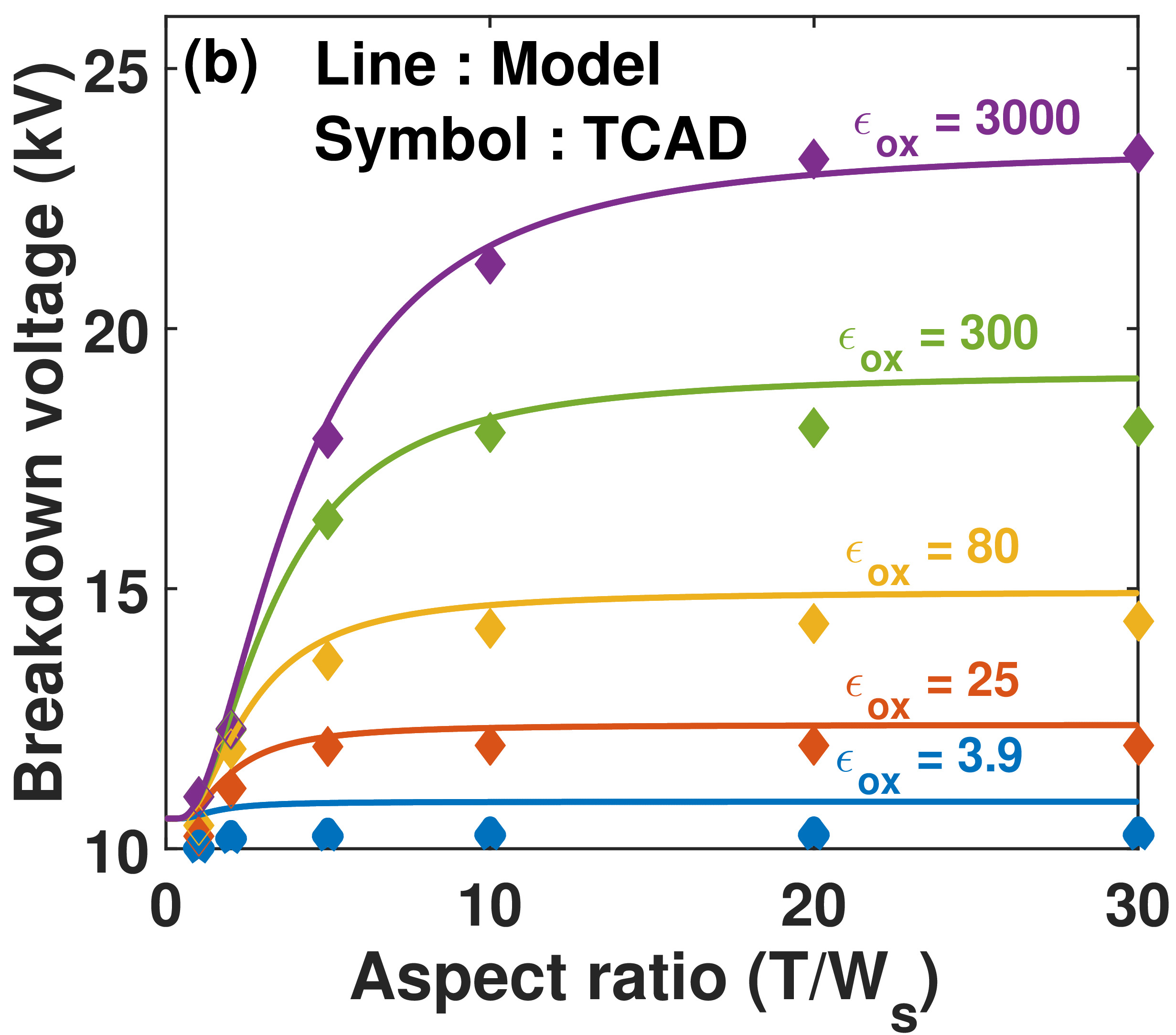}}
\caption{\footnotesize (a) R$_{on}$ vs BV plot with different aspect ratios (b) Variation of breakdown voltage as a function of aspect ratio with the dielectric constant of the dielectric material as the parameter.}
\label{fig7}
\vspace{-0.3cm}
\end{figure}

Fig. 3(a) shows the effect of aspect ratio on the BV-R$_{on}$ trade-off for $\beta$-Ga$_2$O$_3$. It can be observed that as the aspect ratio increases the the BV-R$_{on}$ trade-off is pushed beyond its unipolar limit and the performance saturates after certain aspect ratio (AR $\sim$ 10 ). To find the optimum value of the aspect ratio, breakdown voltage is plotted as a function of aspect ratio for different dielectric constants for an R$_{on}$ of 10 m$\Omega$-cm$^2$ in Fig. 3(b). The appropriate dopings were calculated for the specific R$_{on}$ value using (\ref{eq32}) and the details are presented in the supplemental material. It can be observed that the breakdown voltage saturates at a lower aspect ratio for low-k dielectrics, whereas for high-k dielectrics the aspect ratio can be varied to a much higher value before reaching the maximum achievable breakdown voltage. When a high-k dielectric material is present in contact with an n-type region, the difference in the dielectric permittivity between the semiconductor and the dielectric material results in a larger portion of electric displacement lines flowing through the high-k material. Thus, only a small portion of ionized donors contribute to the electric field in the drift region. Hence with an increase in the dielectric constant, with the R$_{on}$ and aspect ratio kept the same, the lateral electric displacement lines through the dielectric increases , thus increasing the breakdown voltage.

\begin{figure}[t]
\centering
\includegraphics[width=2in, height=7in,keepaspectratio]{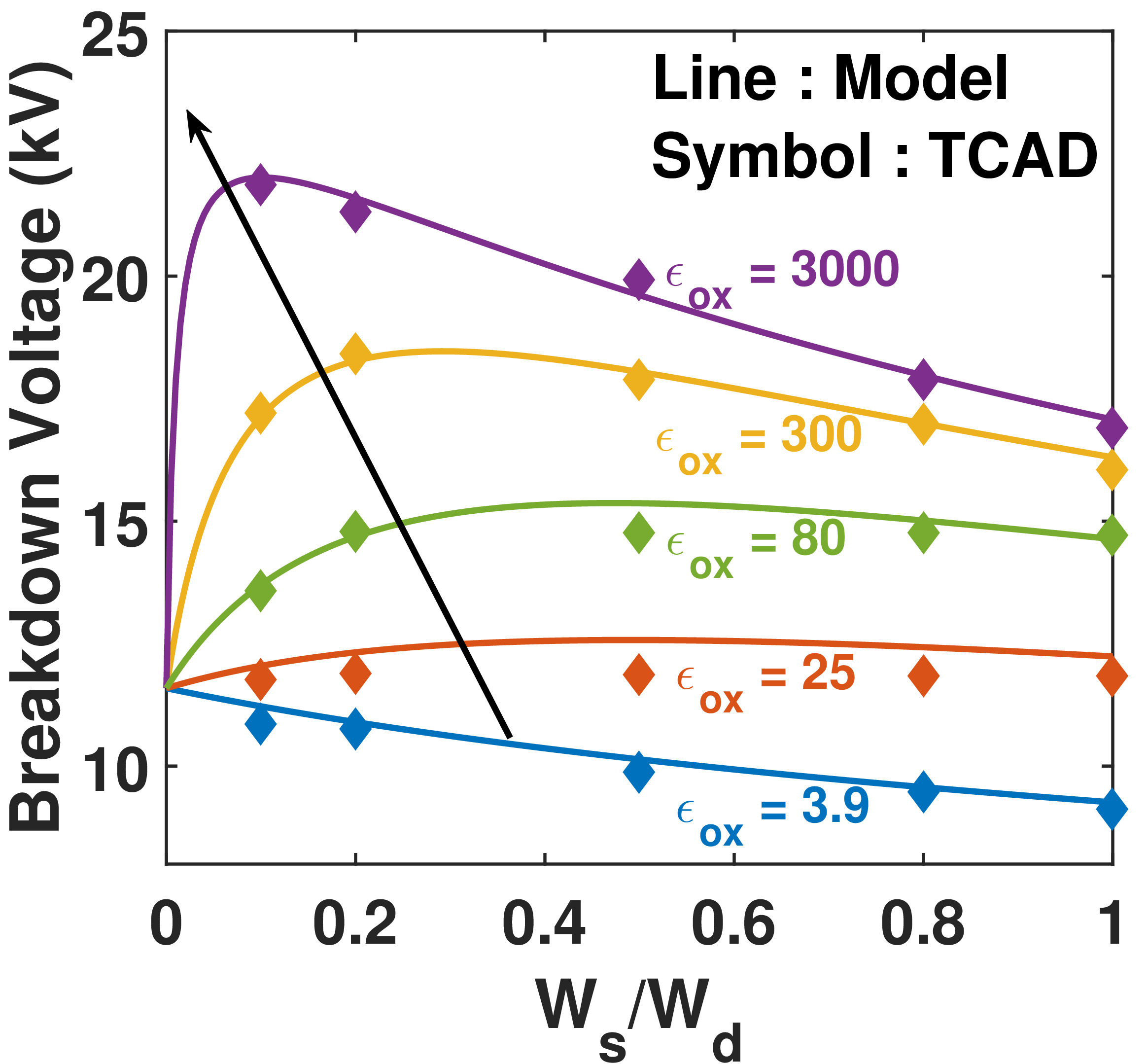}
\caption{\footnotesize Variation of breakdown voltage as a function of semiconductor to dielectric width ratio with the dielectric constant of the dielectric material as the parameter.}
\label{fig6}
\vspace{-0.4cm}
\end{figure}

The effect of the ratio of the semiconductor width to the dielectric width $(W_S/W_D)$ on the breakdown voltage for an R$_{on}$ and aspect ratio of 10 m$\Omega$-cm$^2$ and 10 respectively is plotted in Fig. 4. The optimum breakdown voltage is achieved for a certain $(W_S/W_D)$ for a given dielectric constant of the insulator material. It can be observed that the breakdown voltage first increases as $(W_S/W_D)$ increases, reaches a maximum and then start to decrease with the increase in $(W_S/W_D)$. Also, as the dielectric constant increases, the $(W_S/W_D)$ ratio required for the maximum breakdown voltage reduces. Thus a very narrow  range of $(W_S/W_D)$ is available for extreme permittivity materials to design for the maximum breakdown voltage.

The effect of dielectric constant on the BV-R$_{on}$ trade-off is explored in Fig. 5 (a) and (b) for an aspect ratio of 2 and 10. It is observed that, in the case of a lower aspect ratio (A.R $\sim$ 2), the effect of dielectric constant is very low compared to the one with higher aspect ratio (A.R $\sim$ 10) and if the aspect ratio is less than one the dielectric superjunction has no benefit over the conventional SBDs. The breakdown voltage is also plotted as a function of dielectric constant for an R$_{on}$ of 10 m$\Omega$-cm$^2$ for varying aspect ratios. It can be observed that the breakdown voltage saturates at a lower value of dielectric constant for small aspect ratios, whereas for high aspect ratio, the dielectric constant can be varied to a much higher value before reaching the maximum achievable breakdown voltage. 

\begin{figure}[t]
\centering
\subfigure{
\includegraphics[width=1.71in, height=1.4in,keepaspectratio]{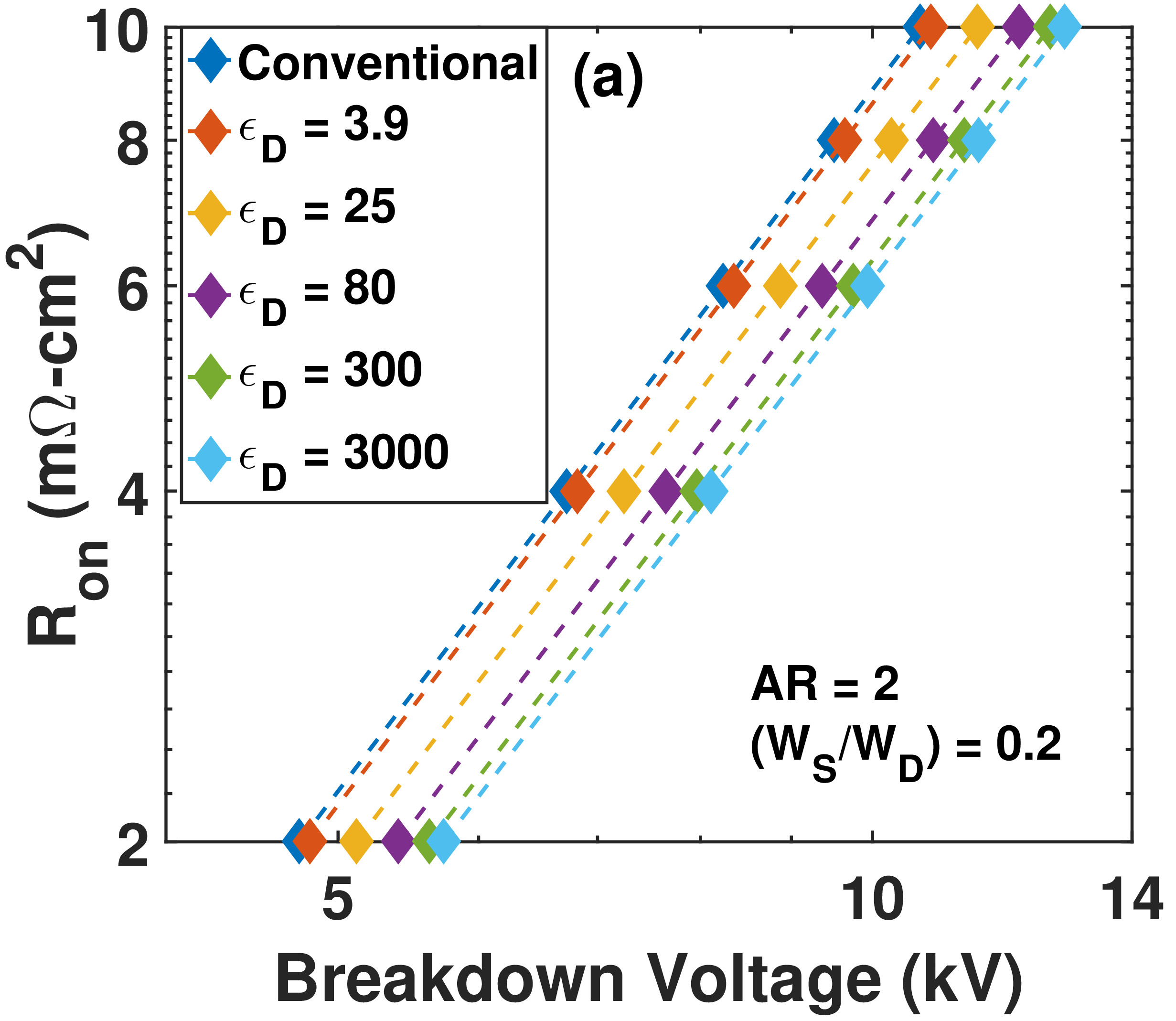}}\hspace{-0.2cm}
\subfigure{
\includegraphics[width=1.71in, height=1.4in,keepaspectratio]{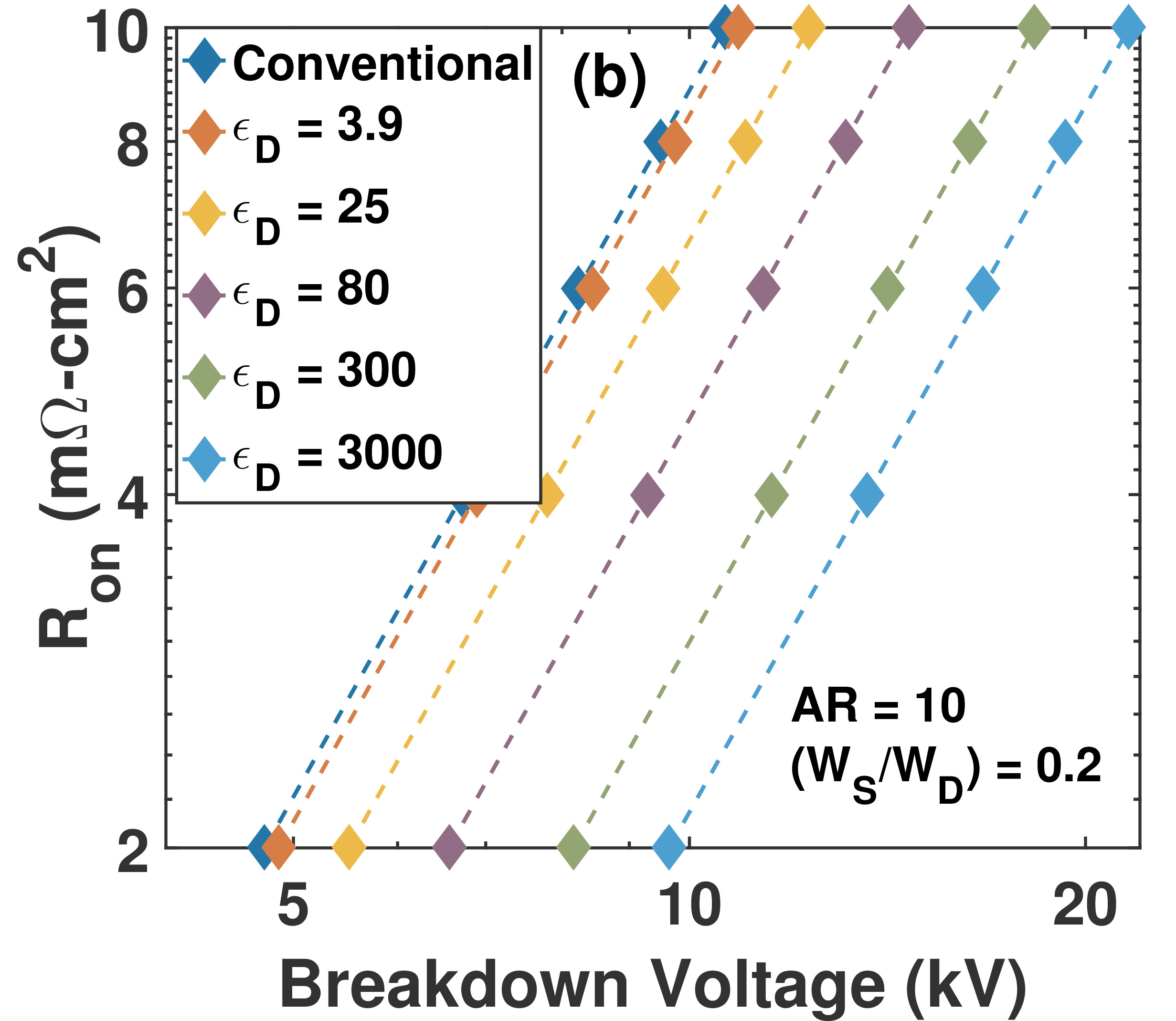}}
\\
\subfigure{
\includegraphics[width=2in, height=3.3in,keepaspectratio]{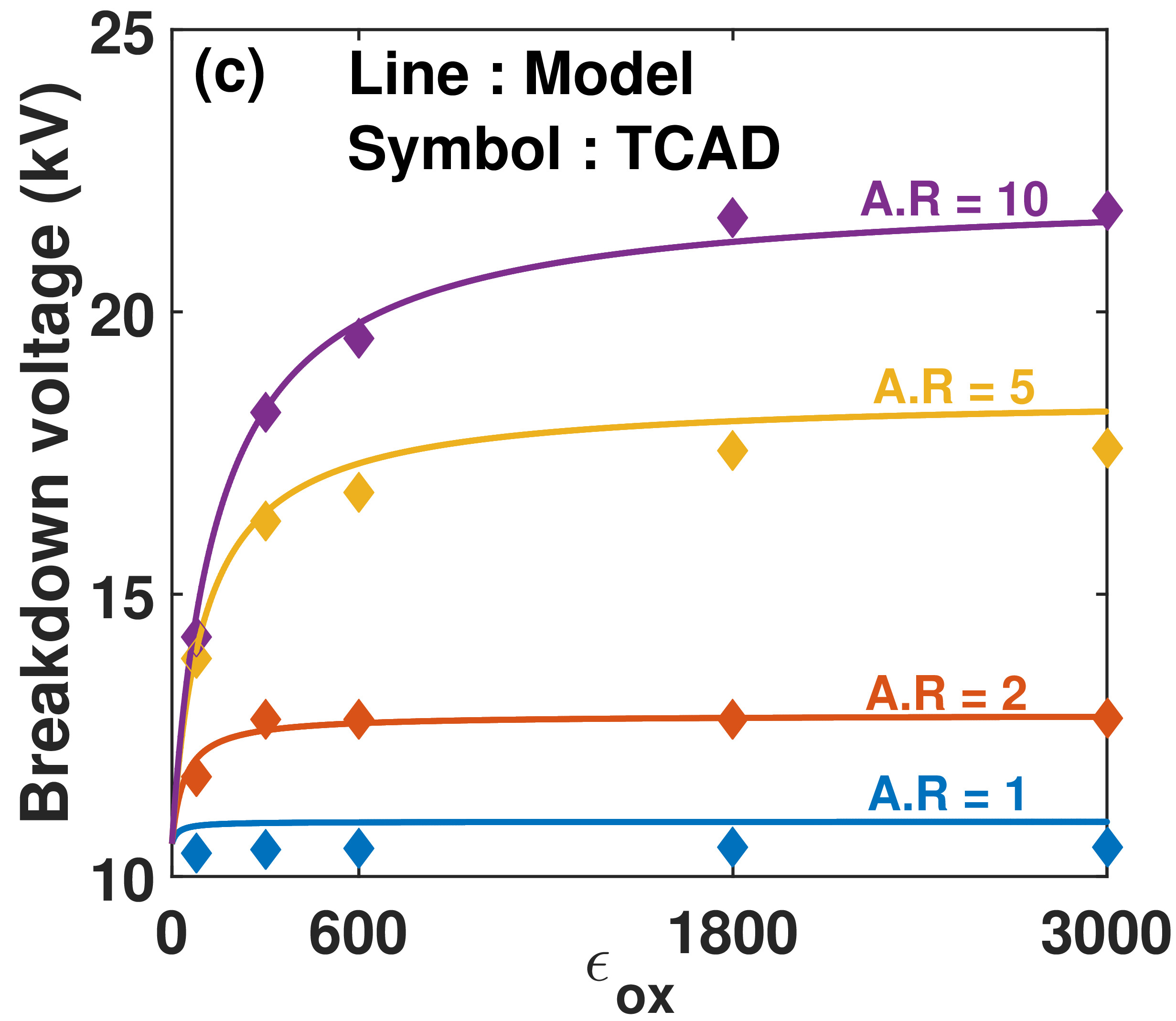}}
\caption{\footnotesize  R$_{on}$ vs BV plot with different dielectric constant of the dielectric material for (a) aspect ratio = 2 and (b) aspect ratio = 2 (c) Variation of breakdown voltage as a function of dielectric constant of the dielectric material with the aspect ratio as the parameter.}
\label{fig7}
\vspace{-0.6cm}
\end{figure}

In summary, the dielectric superjunction SBD using $\beta$-Ga$_2$O$_3$ can be designed with practically achievable device dimensions to achieve extremely high FOM. The aspect ratio and the semiconductor to dielectric width ratio is found to be the most critical parameters as they limit the performance of the structure. If a material with high dielectric constant is used, then the $(W_S/W_D)$ ratio should be adjusted according to Fig. 4 and also according to the design requirement for the R$_{on}$ to achieve the maximum breakdown voltage. A high aspect ratio is also required to achieve better performance as for aspect ratio less than one there would be no benefit of dielectric superjunction SBDs over the conventional SBDs. Thus, a breakdown voltage of 20 kV can be achieved for an R$_{on}$ of 10 m$\Omega$-cm$^2$, dielectric constant of 300, $(W_S/W_D)$ of 0.2 and the aspect ratio of 10, will result in a PFOM of 40 GW/cm$^2$, which is significantly higher than the highest reported FOM in $\beta$-Ga$_2$O$_3$ \cite{huili} and also surpasses the theoretical unipolar figure of merit of $\beta$-Ga$_2$O$_3$ SBD by four times indicating the promise of such structures for exceptional FOM power electronic devices. Further research into the high-k dielectric/$\beta$-Ga$_2$O$_3$ interface will lead to better use of such superjunction structures for ultimate high-performance power electronic devices.

\section*{acknowledgments}
This material is based upon work supported by the Air Force Office of Scientific Research under award number FA9550-18-1-0507 (Program Manager: Dr. Ali Sayir). Any opinions, finding, and conclusions or recommendations expressed in this material are those of the author(s) and do not necessarily reflect the views of the United States Air Force. We also acknowledge funding from the National Science Foundation (NSF) through grant DMR-1931652.


\bibliographystyle{IEEEtran}
\bibliography{main.bib}

\end{document}